\documentstyle[epsfig,twoside,fleqn,espcrc2]{article}
\newcommand{\AmS}{{\protect\the\textfont2
  A\kern-.1667em\lower.5ex\hbox{M}\kern-.125emS}}
\newcommand{\e}{\rm e}
\newcommand{\cO}{{\cal O}}
\newcommand{\cB}{{\cal B}}
\newcommand{\rms}{\rm\scriptsize}

\newcommand{\be}{\begin{equation}}
\newcommand{\ee}{\end{equation}}
\newcommand{\ba}{\begin{array}{c}}
\newcommand{\ea}{\end{array}}
\newcommand{\beqn}{\begin{eqnarray}}
\newcommand{\eeqn}{\end{eqnarray}}

\newcommand{\no}{\nonumber}
\newcommand{\bel}[1]{\be\label{#1}}
 \def\alphas{ \alpha_s }
 \def\alphM{ \alpha_s(M_\tau^2) }

 \def\alphmu{ \alpha_s(\mu^2) }
 
\title{\vspace*{-3.0cm}
\begin{flushright}
{\rm\normalsize FTUV/94-71\\
IFIC/94-68\\ hep-ph/9412273 \\ November 1994\\}
\end{flushright}
\vspace*{0.5cm}
QCD Predictions for the $\tau$ Hadronic Width:
   Determination of $\alpha_s(M_\tau^2)$}
\author{A. Pich\address{Departament de F\'{\i}sica Te\`orica and IFIC,
   Universitat de Val\`encia -- CSIC, \\
   Dr. Moliner 50, E-46100 Burjassot, Val\`encia, Spain}\thanks{Invited talk
   at the QCD 94 Workshop (Montpellier, 7-13 July 1994)}}

\begin{document}

\begin{abstract}
The total $\tau$ hadronic width can be accurately calculated using
analyticity and the operator product expansion.
The theoretical analysis of this
observable is updated to include all available perturbative and
non-perturbative corrections.
The experimental determination of $\alpha_s(M_\tau^2)$ and its
actual uncertainties are discussed.
\end{abstract}


\maketitle

%
%

\section{INTRODUCTION}

The inclusive character of the total $\tau$ hadronic width
renders possible an accurate calculation of the ratio
\cite{BRAATENa,NP,ORSAY,BNP,LDPa,OHIO,NA:94}
%
\be\label{eq:r_tau_def}
     R_\tau \equiv { \Gamma [\tau^- \rightarrow \nu_\tau
                   \,\mbox{\rm hadrons}\, (\gamma)] \over
                         \Gamma [\tau^- \rightarrow
                \nu_\tau e^- {\bar \nu}_e (\gamma)] } ,
\ee
using standard field theoretic methods.
If strong and electroweak radiative corrections are ignored and if
the masses of final-state particles  are neglected,
the universality of the
W coupling to the fermionic charged currents implies
%
\be\label{eq:naive}
R_\tau \, \simeq \, N_c \,
 (|V_{ud}|^2 + |V_{us}|^2) \, \simeq \, 3 \; ,
\ee
 which compares quite well with the  experimental average
 $R_\tau =3.56 \pm 0.03$.
 This provides strong evidence for the colour degree of freedom $N_c$.

 The QCD dynamics is able to account
 quantitatively for the difference between the na\"{\i}ve prediction
(\ref{eq:naive})
 and the measured value of $R_\tau$. Moreover,
 the uncertainties in the theoretical calculation of $R_\tau$ are
 quite small. The value of $R_\tau$ can then be accurately predicted
 as a function of $\alpha_s(M_\tau^2)$.
   Alternatively,  measurements of inclusive $\tau$ decay rates
 can be used to determine the value of the QCD running coupling
 $\alphM$ at the scale of the $\tau$ mass.
 In fact, $\tau$ decay is probably the lowest energy process from which the
 running coupling constant can be extracted cleanly, without hopeless
 complications from non-perturbative effects.
The $\tau$ mass, $M_\tau = 1.7771{\,}^{+0.0004}_{-0.0005}$  GeV \cite{PDG94},
lies fortuitously
 in a ``compromise'' region  where the coupling constant
 $\alphM$ is large enough that $R_\tau$ is sensitive to its
 value, yet still small enough that the perturbative expansion
 still converges well. Moreover,
 the non-perturbative contributions to the total $\tau$-hadronic width are
 very small.

It is the inclusive nature of the total semihadronic decay rate that makes
 a rigorous theoretical calculation of $R_\tau$ possible.
 The only separate contributions to $R_\tau$ that can be calculated are those
 associated with specific quark currents.  We can calculate the
 non-strange and strange contributions to $R_\tau$, and resolve these
 further into  vector and axial-vector contributions.
 Since strange decays cannot be resolved experimentally
into vector and axial-vector  contributions,
 we will decompose our predictions for $R_\tau$
 into only three categories:
\be\label{eq:r_tau_v,a,s}
 R_\tau \, = \, R_{\tau,V} + R_{\tau,A} + R_{\tau,S}\, .
\ee
 Non-strange semihadronic decays of the $\tau$ are resolved experimentally
 into vector ($R_{\tau,V}$) and axial-vector ($R_{\tau,A}$)
 contributions according to whether the
 hadronic final state includes an even or odd number of pions.
 Strange decays ($R_{\tau,S}$) are of course identified by the
 presence of an odd number of kaons in the final state.
 The na\"{\i}ve predictions for these three ratios are
 $R_{\tau,V} \simeq R_{\tau,A} \simeq (N_c/2)|V_{ud}|^2$ and
 $R_{\tau,S} \simeq N_c |V_{us}|^2$, which add up to (\ref{eq:naive}).

\section{THEORETICAL FRAMEWORK}

    The theoretical analysis of $R_\tau$ involves
the two-point correlation functions for
the vector $\, V^{\mu}_{ij} = \bar{\psi}_j \gamma^{\mu} \psi_i \, $
and axial-vector
$\, A^{\mu}_{ij} = \bar{\psi}_j \gamma^{\mu} \gamma_5 \psi_i \,$
colour-singlet quark currents ($i,j=u,d,s$):
\beqn\label{eq:pi_v}
\Pi^{\mu \nu}_{ij,V}(q) &\!\!\!\!\! \equiv &\!\!\!\!\!
 i \int d^4x \, e^{iqx}
\langle 0|T(V^{\mu}_{ij}(x) V^{\nu}_{ij}(0)^\dagger)|0\rangle  ,
\\
\label{eq:pi_a}
\Pi^{\mu \nu}_{ij,A}(q) &\!\!\!\!\! \equiv &\!\!\!\!\!
 i \int d^4x \, e^{iqx}
\langle 0|T(A^{\mu}_{ij}(x) A^{\nu}_{ij}(0)^\dagger)|0\rangle  .
\eeqn
The vector ($V$) and axial-vector ($A$) correlators
have the Lorentz decompositions
\beqn\label{eq:lorentz}
\Pi^{\mu \nu}_{ij,V/A}(q) & \!\!\!\! = & \!\!\!\!
  (-g^{\mu\nu} q^2 + q^{\mu} q^{\nu}) \, \Pi_{ij,V/A}^{(1)}(q^2)   \no\\
  && \!\!\!\! +   q^{\mu} q^{\nu} \, \Pi_{ij,V/A}^{(0)}(q^2) ,
\eeqn
where the superscript $(J=0,1)$ 
denotes the angular momentum in the hadronic rest frame.

The imaginary parts of the two-point functions 
$\, \Pi^{(J)}_{ij,V/A}(q^2) \, $
are proportional to the spectral functions for hadrons with the corresponding
quantum numbers.  The semihadronic decay rate of the $\tau$
can be written as an integral of these spectral functions
over the invariant mass $s$ of the final-state hadrons:
\beqn\label{eq:spectral}
R_\tau  &\!\!\!\!\! = &\!\!\!\!\!
12 \pi \int^{M_\tau^2}_0 {ds \over M_\tau^2 } \,
 \left(1-{s \over M_\tau^2}\right)^2
\no\\ &\!\!\!\!\! \times &\!\!\!\!\!
\biggl[ \left(1 + 2 {s \over M_\tau^2}\right)
 \mbox{\rm Im} \Pi^{(1)}(s)
 + \mbox{\rm Im} \Pi^{(0)}(s) \biggr]  .
\eeqn
 The appropriate combinations of correlators are
\beqn\label{eq:pi}
\Pi^{(J)}(s)  &\!\!\! \equiv  &\!\!\!
  |V_{ud}|^2 \, \left( \Pi^{(J)}_{ud,V}(s) + \Pi^{(J)}_{ud,A}(s) \right)
\no\\ &\!\!\! + &\!\!\!
|V_{us}|^2 \, \left( \Pi^{(J)}_{us,V}(s) + \Pi^{(J)}_{us,A}(s) \right).
\eeqn
 The contributions coming from the first two terms correspond to
$R_{\tau,V}$ and $R_{\tau,A}$ respectively, while
$R_{\tau,S}$ contains the remaining Cabibbo-suppressed contributions.

Since the hadronic spectral functions are sensitive to the non-perturbative
effects of QCD that bind quarks into hadrons, the integrand in
Eq.~(\ref{eq:spectral}) cannot be calculated at present from QCD.
Nevertheless the integral itself can be calculated systematically
by exploiting
the analytic properties of the correlators $\Pi^{(J)}(s)$.
They are analytic
functions of $s$ except along the positive real $s$-axis, where their
imaginary parts have discontinuities.  The integral (\ref{eq:spectral}) can
therefore be expressed as a contour integral
in the complex $s$-plane running
counter-clockwise around the circle $|s|=M_\tau^2$:
\beqn\label{eq:circle}
 R_\tau &\!\!\!\!\! =&\!\!\!\!\!
6 \pi i \oint_{|s|=M_\tau^2} {ds \over M_\tau^2} \,
 \left(1 - {s \over M_\tau^2}\right)^2
\no \\ &\!\!\!\!\!\times &\!\!\!\!\!
 \left[ \left(1 + 2 {s \over M_\tau^2}\right) \Pi^{(0+1)}(s)
         - 2 {s \over M_\tau^2} \Pi^{(0)}(s) \right] \! .
\eeqn

The advantage of expression (\ref{eq:circle})
over (\ref{eq:spectral}) for $R_\tau$
is that it requires the correlators only for
complex $s$ of order $M_\tau^2$, which is significantly larger than the scale
associated with non-perturbative effects in QCD.  The short-distance
Operator Product Expansion (OPE) can therefore be used to organize
the perturbative and non-perturbative contributions
to the correlators into a systematic expansion \cite{SVZ}
in powers of $1/s$,
\be\label{eq:ope}
 \Pi^{(J)}(s) = \sum_{D=2n}\,\sum_{\mbox{\rms dim} {\cal O} = D}
 {{\cal C}^{(J)}(s,\mu) \,\langle {\cal O}(\mu)\rangle\over (-s)^{D/2}} ,
\ee
 where the inner sum is over local gauge-invariant
 scalar operators of dimension $D=0,2,4,\ldots $
 The possible uncertainties associated with the use of the OPE near the
 time-like axis are absent in this case, because
 the integrand in Eq. (\ref{eq:circle}) includes a factor
 $(1- s/M_\tau^2)^2$, which provides a double zero at $s=M_\tau^2$,
 effectively suppressing the contribution from the
 region near the branch cut.
 The parameter $\mu$ in Eq. (\ref{eq:ope})
 is an arbitrary factorization scale, which separates long-distance
 non-perturbative effects, which are absorbed into the vacuum matrix elements
 $\langle {\cal O}(\mu)\rangle $, from short-distance effects, which belong
 in the Wilson coefficients ${\cal C}^{(J)}(s,\mu)$.
 The $D=0$ term (unit operator) corresponds to the pure perturbative
contributions, neglecting quark masses. The leading quark-mass
corrections     generate the $D=2$ term. The first dynamical operators
involving non-perturbative physics appear at $D=4$.
Inserting the functions (\ref{eq:ope})
 into (\ref{eq:circle}) and evaluating the contour integral, $R_\tau$
 can be expressed as an expansion in powers of $1/M_\tau^2$,
 with coefficients that depend only logarithmically on $M_\tau$.

It is convenient to express the corrections to $R_\tau$
from dimension-$D$ operators in terms of the
fractional corrections $\delta^{(D)}_{ij,V/A}$ to the
na\"{\i}ve contribution
from the current with quantum numbers $ij,V$ or $ij,A$:
\beqn\label{eq:r_v}
R_{\tau,V/A} &\!\!\!\!\! = &\!\!\!\!\!  {3 \over 2} |V_{ud}|^2
   S_{EW} \left( 1 + \delta_{EW}' +
      \sum_{D} \delta^{(D)}_{ud,V/A} \right) \! ,
\no \\ \label{eq:r_s} 
R_{\tau,S}\quad
&\!\!\!\!\!\!\!\!\!\!\!\!\!\!\! = &\!\!\!\!\!\!\!\!\!
 3 |V_{us}|^2
   S_{EW} \left( 1 + \delta_{EW}' +
  \sum_{D} \delta^{(D)}_{us} \right) \! .
\eeqn
$\delta^{(D)}_{ij} = (\delta^{(D)}_{ij,V} + \delta^{(D)}_{ij,A})/2$
is the average of the vector 
and axial-vector 
corrections.
 The dimension-0 contribution is the
 purely perturbative correction neglecting quark masses,
 which is the same for all the components of $R_\tau$:
 $\delta^{(0)}_{ij,V/A} = \delta^{(0)}$.
The factors $S_{EW}$ and $\delta_{EW}'$ contain the known electroweak
corrections.
Adding the three terms, the total ratio $R_\tau$ is
\beqn\label{eq:r_total}
R_{\tau} &\!\!\!\!\! = &\!\!\!\!\!
 3 \left( |V_{ud}|^2 + |V_{us}|^2 \right)
S_{EW} \biggl\{ 1 + \delta_{EW}' + \delta^{(0)}
\no\\ &\!\!\!\!\! + &\!\!\!\!\!
\sum_{D=2,4,...}
         \left( \cos^2 \theta_C \delta^{(D)}_{ud}
         + \sin^2 \theta_C \delta^{(D)}_{us} \right) \biggr\} ,
\eeqn
where
$\sin^2\theta_C\equiv |V_{us}|^2/(|V_{ud}|^2 + |V_{us}|^2)$.

\subsection{Perturbative corrections}

In the chiral limit ($m_u=m_d=m_s=0$),
the vector and axial-vector currents are conserved.
This implies  $s \Pi^{(0)}(s) = 0$; therefore, only the correlator
$\Pi^{(0+1)}(s)$
 contributes to Eq.~(\ref{eq:circle}).
Owing \cite{TRUEMAN} to the chiral invariance of massless QCD,
$\Pi^{(0+1)}_{ij,V}(s) = \Pi^{(0+1)}_{ij,A}(s) \equiv \Pi(s)$
 ($i\not=j$) at any finite
order in $\alpha_s$. Moreover, this result is flavour-independent.

The perturbative QCD contribution to $R_\tau$ can then be extracted from
the analogous calculation of the ratio
 $R_{e^+e^-}(s)$ for $e^+e^-$ annihilation,
\beqn\label{eq:r_ee}
 R_{e^+e^-}(s) &\!\!\!\!\equiv&\!\!\!\!
{\sigma(e^+ e^- \rightarrow\,\mbox{\rm hadrons)}
         \over \sigma(e^+e^-\rightarrow\mu^+\mu^-)}
\no\\ &\!\!\!\!
       = &\!\!\!\! 12 \pi \, \mbox{\rm Im} \Pi_{\mbox{\rms em}}(s) \, ,
\eeqn
where $\Pi_{\mbox{\rms em}}(s)$ is the  correlator
associated with the conserved
electromagnetic current
$J^{\mu}_{\mbox{\rms em}}
\equiv \sum_i Q_i \bar{\psi}_i \gamma^{\mu} \psi_i \, \,
(i=u,d,s,...)$.
The result is more conveniently expressed in terms of the
logarithmic derivative of the two-point correlation
function of the vector (axial)
current,
\be\label{eq:d}
D(s)  \equiv  - s {d \over ds } \Pi(s)
=  {1\over 4 \pi^2} \sum_{n=0}  K_n
\left( {\alpha_s(s)\over \pi}\right)^n  ,
\ee
 which satisfies a homogeneous Renormalization Group Equation.
The $K_n$ coefficients are known
to order $\alpha_s^3$ \cite{RUSOS,GKL,SURGULADZE}.
For three flavours, 
one has:
$K_0 = K_1 = 1$; $K_2 = 1.6398$; 
$K_3(\overline{MS}) = 6.3711$.

The perturbative component of $R_\tau$ is given by
\be\label{eq:r_k_exp}
R_\tau^{\mbox{\rms pert}}\equiv 3 \,\{ 1 + \delta^{(0)} \}  =  3
\sum_{n=0}  K_n \, A^{(n)}(\alpha_s) ,
\ee
where
the functions \cite{LDPa}
\beqn\label{eq:a_xi}
A^{(n)}(\alpha_s) &\!\!\!\! = &\!\!\!\! {1\over 2 \pi i}
\oint_{|s| = M_\tau^2} {ds \over s} \,
  \left({\alpha_s(-s)\over\pi}\right)^n
\no\\ &\!\!\!\!\times  &\!\!\!\!
 \left( 1 - 2 {s \over M_\tau^2} + 2 {s^3 \over M_\tau^6}
         - {s^4 \over  M_\tau^8} \right)
\eeqn
are contour integrals
in the complex plane,
which only depend on
$\alphM$.

The running coupling $\alpha_s(-s)$ in Eq. (\ref{eq:a_xi})
can be expanded in powers of
$\alphM$, with coefficients that are polynomials in
$\log{(-s/M_\tau^2)}$.
The perturbative expansion of $\delta^{(0)}$ in powers of
$a_\tau\equiv\alpha_s(M_\tau^2)/\pi$  then takes
\cite{BRAATENa,NP,ORSAY,BNP,LDPa,OHIO}
the form\footnote{
In ref. \protect\cite{BNP} the perturbative contribution to $R_\tau$
was parametrized in terms of the coefficients
$F_n$, appearing in the expansion  of
the spectral function $\mbox{\rm Im}\Pi(s)$ in powers of $\alpha_s(s)/\pi$.
Both parametrizations are related by trivial factors:
$K_2 = F_3$; $K_3 = F_4 + (\pi^2 \beta_1^2/12)$.}
%
$\delta^{(0)} \, = \,\sum_{n=1} (K_n + g_n) \, a_\tau^n$,
where the $g_n$ coefficients depend on $K_{m<n}$
and on $\beta_{m<n}$
[$\beta_1=-9/2$, $\beta_2=-8$,
$\beta_3(\overline{MS})=-3863/192$, ..., are the coefficients of the
QCD $\beta$-function]:
\beqn\label{eq:delta_0}
\delta^{(0)} &\!\!\!\! = &\!\!\!\! a_\tau 
  + \left( K_2 - {19\over 24} \beta_1 \right)
a_\tau^2  
 + \left( K_3 - {19 \over 12} K_2 \beta_1
\right.\nonumber \\ & \!\!\!\! & \!\!\!\!\left.
 - {19 \over 24} \beta_2
         + {265 - 24 \pi^2\over 288} \beta_1^2 \right)
a_\tau^3  
 + \, \cO (a_\tau^4)  \nonumber \\
&\!\!\!\! = &\!\!\!\! a_\tau   
  + 5.2023 \, a_\tau^2  
 + 26.366 \, a_\tau^3  
       + \, \cO(a_\tau^4)  .
\eeqn

One observes  \cite{LDPa} that the
$g_n$ contributions are larger than the direct $K_n$
contributions ($g_2  = 3.563$, $g_3 = 19.99$).
For instance, the bold-guess value
$K_4 \sim K_3 (K_3/K_2)\approx 25$
is to
be compared with $g_4=78$.
These large ``running'' contributions are responsible for the sizeable
renormalization scale dependence found in ref. \cite{CHKL}.
The reason of such uncomfortably large contributions \cite{LDPa} stems  from
the complex integration along the circle $s=M_\tau^2\exp{(i\phi)}$
($\phi\epsilon [0, 2\pi]$) in Eq. (\ref{eq:a_xi}),
which generates the $g_n$ terms.
When the running coupling $\alpha_s(-s)$
is expanded in powers of $\alphM$, one
gets imaginary logarithms, $\log{(-s/M_\tau^2)} = i (\phi - \pi)$,
which are large in some parts of the integration range.
The radius of convergence of this expansion is actually quite small.
A numerical analysis of the series \cite{LDPa}
shows that, at the three-loop level,
an upper estimate for the convergence radius  is
$a_{\tau,\mbox{\rms conv}} \, < 0.11$.

Note\footnote{
A similar suggestion was made in ref.
\protect\cite{PIVOVAROV}.}, however,
that there is no deep reason to stop the  $A^{(n)}(\alpha_s)$ integral
expansions at ${\cal O}(\alphas^3)$.
One can calculate
the $A^{(n)}(\alpha_s)$ expansion  to all orders in
$\alphas$, apart from the unknown
$\beta_{n>3}$ contributions, which are
likely to be
negligible.
Even for $a_\tau$
larger
than the radius of convergence $a_{\tau,\mbox{\rms conv}}$, the integrals
$A^{(n)}(\alpha_s)$ are well-defined functions that can be numerically
computed,
 by using in Eq. (\ref{eq:a_xi}) the exact
solution for $\alpha_s(s)$ obtained from the renormalization-group
$\beta$-function equation.
Thus a more appropriate approach \cite{LDPa} is to
 use a $K_n$ expansion of $R_\tau^{\mbox{\rms pert}}$ as in
Eq.~(\ref{eq:r_k_exp}),
and to fully keep the known three-loop-level calculation of the functions
$A^{(n)}(\alpha_s)$. The perturbative uncertainties are then reduced to the
corrections coming from the unknown $\beta_{n>3}$ and $K_{n>3}$
contributions, since the $g_n$ contributions are properly resummed
to all orders.
To appreciate the size of the effect, Table~\ref{tab:a}
gives the exact results \cite{LDPa}
for $A^{(n)}(\alpha_s)$ ($n=1,2,3$) obtained at the one-, two- and three-loop
approximations (i.e. $\beta_{n>1}=0$, $\beta_{n>2}=0$,
 and $\beta_{n>3}=0$,
respectively), together with the final value of
$\delta^{(0)}$,
for
$a_\tau=0.1$. For comparison, the numbers coming from the truncated
expressions at order $a_\tau^3$ are also given.
Although the difference between the
exact and truncated results represents a tiny
$0.6\% $ effect on $R_\tau$,
it produces a sizeable $4\% $ shift on the value of $\delta^{(0)}$.
The $\delta^{(0)}$ shift, which reflects into a corresponding
shift in the experimental $\alpha_s(M_\tau^2)$ determination,
depends strongly on the value of the coupling constant; for
$a_\tau=0.14$ the $\delta^{(0)}$ shift reaches the $-20\%$ level.

%
\begin{table}[tbh]
\centering
\caption{
Exact results
for $A^{(n)}(\alpha_s)$ ($n=1,2,3$) obtained at the one- ($\beta_{n>1}=0$),
two- ($\beta_{n>2}=0$) and three-loop ($\beta_{n>3}=0$)
approximations, together with the final value of
$\delta^{(0)}$, for
$a_\tau=0.1$. For comparison, the numbers coming from the truncated
expressions at order $a_\tau^3$ are also given}
\hphantom{}
\label{tab:a}
\begin{tabular}{ccccc}
\hline
Loops & $A^{(1)}$ &
$A^{(2)}$ & $A^{(3)}$ &
$\delta^{(0)}$
\\ \hline  
$1$ & $0.13247$ & $0.01570$ & $0.00170$ & $0.1690$ \\
$2$ & $0.13523$ & $0.01575$ & $0.00163$ & $0.1714$ \\
$3$ & $0.13540$ & $0.01565$ & $0.00160$ & $0.1712$ \\
 \hline   
$\cO(\alpha_s^3)$ &
   $0.14394$ & $0.01713$ & $0.00100$ & $0.1784$
%
\\ \hline
\end{tabular}
\end{table}
%

%
\begin{table}[thb]
\centering
\caption{$\delta^{(0)}$ for different values of
$\overline\alpha_s\equiv\alpha_s(M_\tau^2)$}
\hphantom{}
\label{tab:perturbative}
\begin{tabular}{cc|cc}  
\hline
$\overline\alpha_s$ & $\delta^{(0)}$ &
$\overline\alpha_s$ & $\delta^{(0)}$  
\\ \hline
%
$0.24$ & $0.118\pm0.003$   &
$0.34$ & $0.191\pm0.009$
\\
$0.26$ & $0.132\pm0.004$   &
$0.36$ & $0.205\pm0.010$
\\
$0.28$ & $0.146\pm0.005$   &
$0.38$ & $0.220\pm0.012$
\\
$0.30$ & $0.161\pm0.006$   &
$0.40$ & $0.234\pm0.013$
\\
$0.32$ & $0.176\pm0.008$   &
$0.42$ & $0.248\pm0.013$
\\  \hline
\end{tabular}
\end{table}
%

Notice that the difference between using the one- or two-loop approximation to
the $\beta$-function is already quite small
($1.4\%$ effect on  $\delta^{(0)}$), while the change induced by the
three-loop corrections is completely negligible ($0.1\%$). Therefore (unless
the $\beta$-function has some unexpected pathological behaviour at higher
orders), the error induced by the truncation of the $\beta$-function at third
order should be smaller than $0.1\% $ and therefore can be safely neglected.

The only relevant perturbative uncertainties come  from
the unknown higher-order coefficients $K_{n>3}$.
To obtain an estimate of the error induced on $\delta^{(0)}$,
we will take
$\Delta(\delta^{(0)}) \sim \pm K_4 \, A^{(4)}(\alpha_s)$.
The na\"{\i}ve guess
$K_4 \sim (K_3/K_2) K_3 \approx 25$
gives, for
$a_\tau=0.1$, a small $\Delta(\delta^{(0)}) = \pm 0.004$ effect.
The sensitivity on the choice of renormalization scale and
renormalization scheme has been studied in ref. \cite{LDPa}, where it has
been shown to be very small.

The perturbative contribution $\delta^{(0)}$, obtained through
Eqs. (\ref{eq:r_k_exp}) and (\ref{eq:a_xi}), is given in
Table~\ref{tab:perturbative} for different values of
the strong coupling constant $\alpha_s(M_\tau^2)$.
In order to be conservative, and to account for all possible sources
of perturbative uncertainties,
we have  used \cite{OHIO} $K_4 = 50$  for estimating $\Delta(\delta^{(0)})$.

\subsection{Leading quark-mass corrections}

The $1/M_\tau^2$ contributions $\delta^{(2)}_{ij}$ to the ratio $R_\tau$
are simply the leading quark-mass corrections to the perturbative
QCD result of the previous section.
These contributions are known \cite{BNP,CK:93}
to order $\alpha_s^2$.
Quark-mass corrections are certainly tiny for the up and down quarks
($\delta^{(2)}_{ud}\sim -0.08\% $), but
the correction from the strange-quark mass is important
for strange decays \cite{BNP,CK:93}: ($m_u=m_d=0$)
\bel{eq:delta_dos}
\delta^{(2)}_{us,V/A} = -8 {\overline m_s^2\over M_\tau^2}
\left\{1 + {16 \over 3} a_\tau  + 11.03\, a_\tau^2\right\} ,
\ee
where $\overline m_s\equiv m_s(M_\tau^2)$
is the running mass of the strange quark
evaluated at the scale $M_\tau$.
For $\alphM = 0.33$, $\delta^{(2)}_{us}\approx  -19\% $;
nevertheless, because of the $\sin^2{\theta_C}$ suppression, the
effect on the total ratio $R_{\tau}$  is only $-(0.9\pm 0.2) \%$.

\subsection{Non-perturbative contributions}

Since the $\tau$ mass is a quite low energy scale, we should
worry about possible non-perturbative effects.  
In the framework of the OPE, the long-distance dynamics is absorbed
into the vacuum matrix elements
$\langle {\cal O}(\mu)\rangle $, which are (at present) quantities
to be fixed phenomenologically.
If the logarithmic dependence of the Wilson coefficients
${\cal C}^{(J)}(s,\mu)$
on $s$ is neglected
(this is an effect of order $\alpha_s^2$),
the contour integrals
can be evaluated trivially using Cauchy's residue theorem,
and are non-zero only for $D= 2, 4, 6$ and $8$.
The corrections simplify even further if we also
take the chiral limit ($m_u =m_d =m_s =0$).
The dimension-2 corrections then vanish because
there are no operators of dimension 2.  In the chiral limit,
$s \Pi^{(0)}(s) =0$;
thus only the $\Pi^{(0+1)}(s)$ term in
Eq.~(\ref{eq:circle}) contributes to $R_\tau$.
The form of  the kinematical factor multiplying $\Pi^{(0+1)}(s)$
in Eq.~(\ref{eq:circle})  is such that, when the $s$-dependence of the
Wilson coefficients is ignored,
only the $D=6$ and $D=8$ contributions survive the integration.
The power corrections to $R_\tau$  then reduce to \cite{BRAATENa,NP,ORSAY,BNP}
\beqn\label{eq:np_naive}
  \delta^{(6)}_{ij,V/A} &\!\!\! \simeq &\!\!\! - 24 \pi^2
      { \left[ \sum {\cal C}^{(0+1)}_{ij,V/A}
\langle{\cal O}\rangle \right]^{(D=6)}
         \over M_\tau^6}\, ,
\\
  \delta^{(8)}_{ij,V/A} &\!\!\! \simeq &\!\!\! - 16 \pi^2
     { \left[ \sum {\cal C}^{(0+1)}_{ij,V/A}
\langle{\cal O}\rangle \right]^{(D=8)}
         \over M_\tau^8} \, ,
\eeqn
and
$\delta^{(2n)}_{ij,V/A} \simeq 0$   for $2 n \not= 6,8$.

 When the logarithmic dependence
 of the Wilson coefficients on $s$ is taken into account, operators of
 dimensions other than 6 and 8 do
 contribute, but they are suppressed by two powers
 of $\alphM$.
 The largest power corrections to $R_\tau$  then come from
 dimension-6 operators, which have no such suppression.
  	Their size  has been estimated
in ref. \cite{BNP}, using published phenomenological fits to different
sets of data:
\be\label{eq:delta_6}
\delta^{(6)}_{ij} \approx -{2\over 7}
\delta^{(6)}_{ij,V} \approx {2\over 11}
\delta^{(6)}_{ij,A} \approx -(0.7\pm0.4)\% .
\ee
These power corrections are numerically
very small, which is due to the fact that they fall off
like the sixth power of $1/M_\tau$.
Moreover, there is a large cancellation
between the vector and axial-vector contributions
to the total hadronic width
(the operator with the largest Wilson coefficient contributes with
opposite signs to the vector and axial-vector correlators, due to the
$\gamma_5$ flip). Thus, the
non-perturbative corrections to $R_\tau$ are smaller than the
corresponding contributions to  $R_{\tau,V/A}$.
A more detailed study of non-perturbative corrections, including
the very small $D=4$ contributions proportional to quark masses
or to $\alphM^2$, can be found in ref. \cite{BNP}.

\subsection{Electroweak corrections}

The electroweak corrections to the ratio $R_\tau$
are quite sizeable, because the
corrections to the numerator 
include
logarithms of $M_Z/M_\tau$, which are not present in the corrections
to the denominator. These logarithms represent a short-distance correction
to the low-energy effective
four-fermion coupling of the $\tau$ to $\nu_\tau d {\bar u}$
or $\nu_\tau s {\bar u}$, and, therefore, should
be absorbed into an overall multiplicative factor $S_{EW}$.
Using the renormalization group to sum up higher-order
$\alpha^n \log^n(M_Z/M_\tau)$ contributions,
$S_{EW}$ becomes \cite{MS}
\beqn\label{eq:s_ew}
 S_{EW} &\!\!\!\!\! = &\!\!\!\!\!
\left( {\alpha(m_b^2) \over \alpha(M_\tau^2)} \right)^{9\over19}
         \left( {\alpha(M_W^2) \over \alpha(m_b^2)} \right)^{9\over 20}
         \left( {\alpha(M_Z^2) \over \alpha(M_W^2)} \right)^{36\over 17}
\no\\ &\!\!\!\!\! = &\!\!\!\!\!
         1.0194 \, .
\eeqn
The residual non-logarithmic electroweak correction
is quite small \cite{BLI}:
\be\label{eq:d_ew'}
 \delta_{EW}' = (5/12) \, \alpha(M_\tau^2)/\pi \simeq 0.0010 .
\ee

\section{DETERMINATION OF $\alpha_s(M_\tau^2)$}

%
\begin{table}[t] 
\centering
\caption{QCD predictions for the different components
 of the $\tau$ hadronic width as function of
$\overline\alpha_s\equiv\alpha_s(M_\tau^2)$}
\hphantom{}
\label{tab:predictions}
\begin{tabular}{cccc}
\hline
$\overline\alpha_s$ & $R_{\tau,V}$     & $R_{\tau,A}$ & $R_{\tau,S}$
\\ \hline
$0.24$   & $1.66 \pm 0.02$  & $1.56 \pm 0.03$ & $0.145\pm 0.005$
\\
$0.26$   & $1.68 \pm 0.02$  & $1.58 \pm 0.03$ & $0.145\pm 0.005$
\\
$0.28$   & $1.70 \pm 0.02$  & $1.61 \pm 0.03$ & $0.145\pm 0.005$
\\
$0.30$   & $1.72 \pm 0.02$  & $1.63 \pm 0.03$ & $0.145\pm 0.006$
\\
$0.32$   & $1.75 \pm 0.02$  & $1.65 \pm 0.03$ & $0.145\pm 0.006$
\\
$0.34$   & $1.77 \pm 0.02$  & $1.67 \pm 0.03$ & $0.145\pm 0.006$
\\
$0.36$   & $1.79 \pm 0.02$  & $1.69 \pm 0.03$ & $0.144\pm 0.006$
\\
$0.38$   & $1.81 \pm 0.03$  & $1.71 \pm 0.03$ & $0.144\pm 0.007$
\\
$0.40$   & $1.83 \pm 0.03$  & $1.73 \pm 0.03$ & $0.143\pm 0.007$
\\
$0.42$   & $1.85 \pm 0.03$  & $1.75 \pm 0.04$ & $0.143\pm 0.007$
\\ \hline
\end{tabular}
\end{table}
%

The final QCD predictions \cite{BNP,LDPa,OHIO} for
$R_{\tau,V}$, $R_{\tau,A}$, $R_{\tau,S}$ and
$R_{\tau}$ are given in Tables~\ref{tab:predictions} and \ref{tab:rtau},
as functions of the coupling constant
$\alpha_s(M_\tau^2)$.

The experimental value for $R_\tau$ is actually
determined by measuring the leptonic branching fractions:
\be\label{eq:rtauexpb}
     R_\tau^{B} \equiv { 1 - B_e - B_\mu \over B_e } \, ,
\ee
where $B_\ell = \Gamma[\tau^- \rightarrow \nu_{\tau}
           \ell^- \bar{\nu}_\ell (\gamma)] / \Gamma_\tau$
and $\Gamma_\tau$ is the total decay rate.

An independent determination of $R_\tau$ can be obtained by measuring the
lifetime.
Because the decays
$\tau^- \rightarrow \nu_\tau   \ell^- \bar{\nu}_\ell (\gamma)$
are purely electroweak processes, their rates
$\Gamma _{\tau \rightarrow \ell} = B_l/\tau_\tau$ can be calculated
theoretically with great accuracy.
The only unknown in Eq. (\ref{eq:rtauexpb}) is therefore
the total decay rate
\cite{PI:93}:
\be\label{eq:rtauexpg}
     R_\tau^{\Gamma} =
{(1.632\pm 0.002)\times 10^{-12}\,\mbox{\rm s}\over \tau_\tau} - 1.97257 .
\ee
%

%
\begin{table}[t] 
\centering
\caption{$R_\tau$ for different values of
$\overline\alpha_s\equiv\alpha_s(M_\tau^2)$}
\hphantom{}
\label{tab:rtau}
\begin{tabular}{cc|cc}
\hline
$\overline\alpha_s$ & $R_\tau$ & $\overline\alpha_s$ & $R_\tau$
\\ \hline
$0.24$ & $3.37 \pm 0.02$   &
$0.34$ & $3.58 \pm 0.03$
\\
$0.26$ & $3.41 \pm 0.02$   &
$0.36$ & $3.63 \pm 0.03$
\\
$0.28$ & $3.45 \pm 0.02$   &
$0.38$ & $3.67 \pm 0.04$
\\
$0.30$ & $3.50 \pm 0.02$   &
$0.40$ & $3.71 \pm 0.04$
\\
$0.32$ & $3.54 \pm 0.03$   &
$0.42$ & $3.75 \pm 0.04$
\\  \hline
\end{tabular}
\end{table}
%

The present \cite{PDG94} results for these two independent determinations
of $R_\tau$ are
\beqn\label{rexp}
 R_\tau^{B} &\!\!\! =&\!\!\! 3.56 \pm 0.04 \, ,\\
 R_\tau^{\Gamma} &\!\!\! =&\!\!\! 3.55 \pm 0.06 \, .
\eeqn
Using the predictions in Table~\ref{tab:rtau},
the average of the two experimental determinations of $R_\tau$,
\be
  R_\tau =3.56 \pm 0.03 \, ,
\ee
corresponds to
\be\label{eq:alpha}
\alpha_s(M_\tau^2)  =  0.33\pm0.03 \, .
\ee

Once the running coupling constant $\alphmu$ is determined at the scale
$M_\tau$, it can be evolved to higher energies using the renormalization
group.  The error bar on $\alphmu$ must also be evolved using
the renormalization group.  Its size scales roughly
as $\alphmu^2$, and it therefore shrinks as $\mu$ increases.
Thus a modest precision in the
determination of $\alpha_s$ at low energies results in a very high
precision in the coupling constant at high energies.
After evolution up to the scale $M_Z$, the strong coupling constant in
Eq. (\ref{eq:alpha}) decreases to
\be\label{eq:alpha_z}
\alpha_s(M_Z^2)  =  0.120{\,}^{+0.003}_{-0.004} \, ,
\ee
%
in excellent
agreement with the present LEP average (without $R_\tau$)
$\alpha_s(M_Z^2)  =  0.123\pm0.006$ \cite{LEP}, and with a similar error bar.
The comparison of these two determinations of $\alphmu$ in two extreme
energy regimes, $M_\tau$ and $M_Z$, provides a beautiful test of the
predicted running of the QCD coupling constant.

\subsection{Semi-inclusive $\tau$-decay widhts}

With $\alphM$ fixed to the value in Eq.~(\ref{eq:alpha}),
Table~\ref{tab:predictions} gives definite
predictions for the semi-inclusive $\tau$-decay widths:
$R_{\tau,V}= 1.76\pm 0.06$,
$R_{\tau,A}= 1.66\pm 0.06$,
$R_{\tau,S}= 0.145\pm 0.006$.
The experimental  results on exclusive hadronic
$\tau$-decay modes provide then a consistency check of the reliability
of the theoretical analysis.

The assignment of a given measurement to one of the three categories
$R_{\tau,V}$, $R_{\tau,A}$, $R_{\tau,S}$, is not completely straightforward.
One needs to have a clean $\pi / K$ identification, and to
know the exact number of neutral particles, to separate
the vector and axial-vector contributions.
Fortunately, the improved quality of the recent data has allowed to
perform an explicit identification of multiple $\pi^0$'s
\cite{neutrals,NEWneutrals}
and the first systematic analysis \cite{kaons}
of exclusive Cabibbo-suppressed decays.

The Particle Data Group \cite{PDG94} has made a constrained fit
to the data,
using 
12 basis modes whose branching fractions sum exactly to unity:
$e^-\bar\nu_e\nu_\tau$, $\mu^-\bar\nu_\mu\nu_\tau$, $\pi^-\nu_\tau$,
$K^-\nu_\tau$, $K^*(892)^-\nu_\tau$, $\pi^-\pi^0\nu_\tau$,
$h^-2\pi^0\nu_\tau$, $h^-3\pi^0\nu_\tau$, $h^-4\pi^0\nu_\tau$,
$2h^-h^+\nu_\tau$, $2h^-h^+(\geq 1 h^0)\nu_\tau$ and
$3h^-2h^+(\geq 0 h^0)\nu_\tau$,
where $h^\pm$ and $h^0$ denote charged and neutral hadrons, respectively.
Assigning the modes with an unknown number of neutral pions
to the category corresponding to the minimum
possible multiplicity
[i.e. the decay into
$2h^-h^+(\geq 1 h^0)\nu_\tau$
to $R_{\tau,V}$ and $3h^-2h^+(\geq 0 h^0)\nu_\tau$ to
$R_{\tau,A}$],
this fit provides a first approximation to the actual semi-inclusive
decay widths.
The resulting values can then be slightly corrected using the more
exclusive measurements of the modes
$3h^-2h^+\pi^0\nu_\tau$, $2h^-h^+2\pi^0\nu_\tau$, $\eta\pi^-\pi^0\nu_\tau$,
$\bar K^*(892)^0\pi^-(\geq 0 h^0)\nu_\tau$,
$K^-K^+\pi^-(\geq 0 h^0)\nu_\tau$
and $K^*(892)^-(\geq 0 h^0)\nu_\tau$.
Following this procedure and taking into account the most recent data
\cite{NEWneutrals,kaons,GH:94},
I get\footnote{
The direct sum of the measured exclusive modes into kaons
\protect\cite{kaons,GH:94} gives $R_{\tau,S} = 0.16\pm 0.02$.
}:
\beqn\label{eq:exp_r_v}
R_{\tau,V} &\!\!\!\! =  &\!\!\!\! 1.76\pm 0.04 \, ,
\no\\
R_{\tau,A} &\!\!\!\! =  &\!\!\!\! 1.66\pm 0.05 \, ,
\\
R_{\tau,S} &\!\!\!\! =  &\!\!\!\! 0.14\pm 0.02 \, , \no
\eeqn
which add to $R_\tau = 3.56\pm 0.07$.
The agreement with the theoretical predictions is excellent.

The $R_{\tau,S}$ predictions
are very sensitive to the power corrections. As shown in
Table \ref{tab:predictions},
there is practically no dependence on the value
of the strong coupling in this case.
In fact, the final predictions turn out to be very close to the na\"{\i}ve
expectation $R_{\tau,S}\simeq N_c |V_{us}|^2 \simeq 0.147$,
because there is a strong cancellation between the perturbative contribution
$\delta^{(0)}$ and the strange-quark-mass correction $\delta^{(2)}$.
The success of the theoretical prediction can then be taken in
this case as a test of the $D=2$ contribution.
With more precise data, it could be possible to use $R_{\tau,S}$
to check the value of the strange-quark mass.

\subsection{$e^+e^-$ test}

In the vector channel, one can also use the information obtained,
through an isospin rotation, from the isovector part of the $e^+ e^-$
annihilation cross-section into hadrons. The exclusive $\tau$-decay
width into Cabibbo-allowed modes with $J^P=1^-$ can be  expressed
as an integral over the corresponding $e^+e^-$ cross-section
\cite{TSAI},
\beqn\label{eq:cvc}
R_{\tau,V}  &\!\!\!\! = &\!\!\!\!
   {3 \cos^2{\theta_C} \over 2 \pi \alpha^2 M_\tau^8 } S_{EW}
   \int_0^{M_\tau^2}  ds \; (M_\tau^2 - s)^2
\no\\ &\!\!\!\! &\!\!\!\! \quad\times\,
(M_\tau^2 + 2 s) \, s \,
    \sum_{V^0} \sigma^{I=1}_{e^+ e^- \to V^0}(s) .
\eeqn
The analysis of the separate exclusive vector modes \cite{NP:93,EI}
shows a good agreement between the actual
$\tau$-decay measurements and the numbers obtained from $e^+e^-$ data.
However, the $\tau$-decay data is already more accurate than the
$e^+e^-$ results; therefore, this exercise does not improve the
$R_{\tau,V}$ measurement in Eq.~(\ref{eq:exp_r_v}).
Nevertheless, the $e^+e^-$ data offers us the opportunity to make an
additional test of the QCD predictions, by varying
the value of $M \equiv M_\tau$ in Eq.~(\ref{eq:cvc}).

\begin{figure}[tb]
\epsfig{file=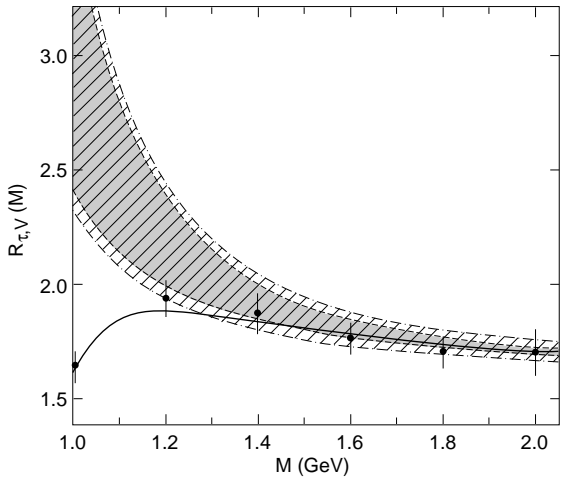,width=7.5cm}
\vspace{-1.1cm}
\caption{$R_{\tau,V}$ as a function of $M\equiv M_\tau$.}
\label{fig:r_tau_m}
\end{figure}

The theoretical predictions for $R_{\tau,V}$ as a function of $M$
can be trivially obtained from the formulae given in refs. \cite{BNP,LDPa}.
For lower values of $M$ it is important to use the
resummed perturbative expansion of ref. \cite{LDPa}, since
the bigger values of $\alpha_s(M^2)$ in that region   imply that
the non-resummed expansion is non-convergent.
Figure~\ref{fig:r_tau_m} \cite{NP:93}
compares the theoretical predictions with the
results obtained from $e^+e^-$ data (the points with vertical error bars).
Above 2 GeV the data is rather conflicting and incomplete, and one
needs to rely on extrapolations of the fits done at lower energies; so
this region has not been plotted\footnote{
Although one finds also quite good agreement between theory and
experiment above 2 GeV, the big error bars of the ``experimental'' points
make the comparison quite meaningless.}.
The shaded area between the two dashed curves
corresponds to the theoretical prediction
for
$\alpha_s(M_\tau^2) = 0.33$. The big allowed region at low values
of $M$ is due to the uncertainty in the leading non-perturbative
correction, taken from the estimate in Eq.~(\ref{eq:delta_6}), i.e.
\be\label{eq:delta_6_M}
\delta^{(6)}_V(M)
= (0.024\pm0.013)\times \left({M_\tau/ M}\right)^6 \, .
\ee
At the $\tau$-mass scale, $\delta^{(6)}_V(M_\tau) = \delta^{(6)}_V$
is a very tiny correction;
however, since it scales as the sixth power of $M$, this non-perturbative
contribution (and its associated uncertainty) increases very
rapidly as $M$ decreases.

%
%
%
%

Allowing the value of $\alpha_s(M_\tau^2)$ to change in the range
(\ref{eq:alpha}),
one gets the larger region between the two dot-dashed curves in
Figure~\ref{fig:r_tau_m}.
For $M \sim M_\tau$, the theoretical uncertainty is dominated by the
uncertainty in the input value of $\alpha_s(M^2_\tau)$;
the error on  $\delta^{(6)}_V$ becomes dominant for $M < 1.6$ GeV,
and overwhelms the result for $M <$ 1.2 GeV.

One can notice that there is a good agreement between the QCD
predictions and the $e^+ e^-$ data points for $M\ge 1.2$ GeV.
This confirms the role of the threshold factor $(1-s/M^2)^2$,
which minimizes the theoretical uncertainties near the physical cut,
and further supports the 
theoretical framework used to
analyze the
$\tau$ hadronic width.

The departure of the theoretical prediction from the data points below
1.2 GeV signals the important role of higher-order power corrections in
this region. The subleading dimension-8 correction has been neglected
before because,  at the $\tau$-mass scale, its contribution
is expected to be smaller than the
uncertainty on $\delta^{(6)}_V$. However,
when going to smaller values of $M$, the $\delta^{(8)}_V(M)$
contribution increases much
faster than the dimension-6 one and at some point would even become
dominant, indicating a breakdown of the  expansion in powers of $1/M$.
We can use the lower-mass data points to make an
estimate of the size of this contribution.
Taking $\alpha_s(M_\tau^2)=0.33$ and $\delta^{(6)}_V = 0.024$,
a quite reasonable fit is obtained for
$\delta^{(8)}_V = -0.0095$.
This is shown by the continuous curve in Figure~\ref{fig:r_tau_m}.
Although the $\delta^{(8)}_V(M)$ correction is tiny at $M=M_\tau$, its
effect changes completely the predicted behaviour below 1.2 GeV.
Note, however, that for this value of $\delta^{(8)}_V$
one has
$\delta^{(8)}_V(M)/\delta^{(6)}_V(M) = -1.25$ at $M = 1$ GeV,
which puts some doubts on the applicability of the inverse power expansion
at such a low scale. If one  takes only into account the region above
1.2 GeV, the size of the experimental error bars does not allow us to make a
clear statement about the size of  $\delta^{(8)}_V$
($\delta^{(8)}_V = 0$ is  compatible with the data),
although smaller values of $\delta^{(8)}_V$ seem to be preferred.

\subsection{Hadronic invariant-mass distribution}

The leading non-perturbative contributions to $R_\tau$ are suppressed
by a factor of $1/M_\tau^6$ and, therefore, are very small.
Nevertheless they introduce a small uncertainty in the predictions,
since their actual evaluation involves a mixture of experimental
measurements and theoretical considerations, which are model-dependent
to some extent.
It would be better to directly measure this contribution from the
$\tau$-decay data themselves. This information can be extracted
\cite{PICHSLAC} from
the invariant-mass distribution of the final hadrons in $\tau$ decay.

Although the distributions themselves cannot be predicted at present,
certain weighted integrals of the hadronic spectral functions can be
calculated in the same way as $R_\tau$.
The analyticity properties of $\Pi^{(J)}_{ij,V/A}$ imply \cite{PICHSLAC,LDPb}:
%
\beqn\label{eq:weighted_integrals}
\int_0^{s_0} ds\, W(s)\, \mbox{\rm Im}\Pi^{(J)}_{ij,V/A} =
\qquad\qquad\qquad && \no\\ \qquad\qquad\qquad
{i\over 2} \oint_{|s|=s_0} ds\, W(s) \,\Pi^{(J)}_{ij,V/A} , &&
\eeqn
with $W(s)$ an arbitrary weight function without singularities in the
region $|s|\leq s_0$.
Generally speaking, the accuracy of the theoretical predictions can be
much worse than the one of $R_\tau$, because non-perturbative effects
are not necessarily suppressed.
In fact, choosing an appropriate weight function, non-perturbative effects
can even be made to dominate the final result. But this is precisely
what makes these integrals interesting: they can be used to measure the
parameters characterizing the non-perturbative dynamics.

To perform an experimental analysis, it is convenient to use
moments of the directly measured invariant-mass distribution
\cite{LDPb} ($k,l\ge 0$)
\be\label{eq:moments}
R^{kl}_\tau(s_0) \equiv\int_0^{s_0}\, ds
 \, \left(1 - {s\over s_0}\right)^k \left ( {s \over M_\tau^2} \right )^l
{ d R_\tau \over d s} .
\ee
The factor $(1-s/s_0)^k$ supplements $(1-s/M_\tau^2)^2$ for $s_0\not=
M_\tau^2$,
in order to squeeze the integrand at the crossing of the positive real-axis
and, therefore, improves the reliability of the OPE analysis; moreover, for
$s_0=M_\tau^2$ it reduces the contribution from the tail of the distribution,
which is badly defined experimentally.
A combined fit of
different $R_\tau^{kl}(s_0)$ moments
results in experimental
values for $\alphM$ and for the coefficients of the inverse-power corrections
in the OPE.

The first experimental study has been done by ALEPH \cite{aleph},
using the moments (0,0), (1,0), (1,1), (1,2) and (1,3).
$R^{00} = R_\tau$ uses the overall normalization of the hadronic
distribution, while the ratios $D_\tau^{kl} = R^{kl}_\tau/R_\tau$ are based on
the shape of the $s$ distribution and are more dependent on
non-perturbative effects \cite{LDPb}.
The resulting strong coupling constant measurement \cite{aleph},
$\alpha_s(M_\tau^2) = 0.330\pm 0.046$,
is in excellent agreement with the $R_\tau$ determination
in Eq.~(\ref{eq:alpha}).
Moreover, the ALEPH analysis has determined the total non-perturbative
contribution to $R_\tau$ to be \cite{aleph}
\bel{eq:delta_NP}
\delta_{\mbox{\rms NP}} = (0.3\pm 0.5)\% ,
\ee
confirming the predicted \cite{BNP}
suppression of non-perturbative corrections.

More recently, ALEPH \cite{DU:94} and CLEO \cite{cleo} have
reported preliminary results of two independent
analyses, with a much larger data sample.
Although their measured $s$-distributions are in excellent agreement
\cite{DU:94}, the fitted $\alpha_s$ values, shown in Table~\ref{tab:moments},
disagree by about 2 standard deviations.
The origin of the discrepancy stems from their different normalizations.
While CLEO uses the world-averaged $B_l$ values quoted by
the Particle Data Group \cite{PDG94}, ALEPH has smaller (and precise)
leptonic branching ratios and therefore a larger $\tau$ hadronic width,
$R_\tau = 3.645\pm 0.024$.
The average of the ALEPH and CLEO $\alpha_s(M_\tau^2)$ determinations,
\bel{eq:alpha_moments}
\alpha_s(M_\tau^2) = 0.335 \pm 0.016 \, ,
\ee
nicely reproduces the result (\ref{eq:alpha}), obtained from $R_\tau$
using the theoretical estimates of the small non-perturbative corrections.
Nevertheless, the error in (\ref{eq:alpha_moments}) should probably be
increased by a factor of 2,
to account for the present $2\sigma$ discrepancy between the
individual measurements.

In addition to confirm that the total non-perturbative correction to $R_\tau$
is indeed small,
the moment analysis allows to make a determination of the $D=4$
gluon condensate (which does not play any role in $R_\tau$), in good
agreement with
the previous phenomenological estimate
$\langle {\alpha_s\over\pi} GG\rangle = 0.02\pm 0.01$ GeV$^4$
\cite{NP,BNP}.

%
\begin{table}[t] 
\centering
\caption{Fitted values of $\alpha_s(M_\tau^2)$ and the
non-perturbative corrections. $\langle {\alpha_s\over\pi} GG\rangle$
is given in units of GeV$^4$
}
\hphantom{}
\label{tab:moments}
\begin{tabular}{ccc}
\hline
& ALEPH & CLEO
\\ \hline
$\alpha_s(M_\tau^2)$ & $0.355\pm 0.021$ & $0.309\pm 0.024$
\\
$\langle {\alpha_s\over\pi} GG\rangle$  &
$0.006\pm 0.012$ & $0.024\pm 0.010$
\\
$\phantom{-}\delta^{(6)}$ & $0.006\pm 0.006$ & $0.014\pm 0.004$
\\
$-\delta^{(8)}$ & $0.0017\pm 0.0014$ & $0.0028\pm 0.0006$
\\ \hline
$\chi^2/\mbox{\rm d.o.f.}$ & 0.002 & 2.1
\\ \hline
\end{tabular}
\end{table}
%

\section{THEORETICAL UNCERTAINTIES}

The accuracy of the experimental data is obviously going to improve
in the near future. Thus, the final error in the determination of
$\alpha_s(M_\tau^2)$
will be limited by the accuracy of the theoretical calculation.
It is then important to carefully study the possible sources of
theoretical uncertainties and try to pin down their numerical effects.

\subsection{Higher-order perturbative corrections}

The precision of the perturbative calculation of $R_\tau$ has been
extensively discussed in ref. \cite{LDPa}. Once the important higher-order
logarithmic effects have been properly identified and resummed to all
orders, the size of the remaining renormalization scheme (and scale)
dependence is quite small. The only relevant perturbative
uncertainties come from the unknown higher-order coefficients $K_{n>3}$.
We have estimated their numerical effect by assuming an algebraic growth
of these coefficients and, moreover, in order to be conservative,
we have further increased this
estimate by a factor of two \cite{OHIO},
i.e. $K_4\sim 2 (K_3/K_2) K_3\approx 50$:
\bel{eq:pert_error}
\Delta(\delta^{(0)}) = \pm K_4\, A^{(4)}(\alpha_s) = \pm 0.008 \,  .
\ee
Clearly, this is an arbitrary prescription,
but it is probably the most conservative one in the absence of a
direct calculation of $K_4$.
Note, that this value of $K_4$
corresponds already to a non-convergent perturbative series:
\bel{eq:divergent_series}
K_3\, A^{(3)}(\alpha_s) = 0.010 \sim
K_4\, A^{(4)}(\alpha_s) .
\ee
A recent estimate of $K_4$, based on the application of scheme-invariant
methods such as the principle of minimal sensitivity \cite{ST:81}
or the effective charges approach \cite{GR:80},
finds \cite{KS:94} $K_4=27.46$.
A similar value $K_4=29\pm 4 \pm 2$ has been obtained from a direct fit to
the experimental data \cite{LD:94}.
The two estimates are indeed smaller than our conservative choice $K_4=50$.

\subsection{Renormalons}

 Possible uncertainties related to
the asymptotic (at best)
nature of the QCD perturbative expansions at large orders have been
considered recently
\cite{WEST,MUELLER,GR:94,DA:84,BROWN,BENEKE,LTM:94,ZAKHAROV,ALTARELLI,AL:94,BZ}.
The leading large-order contributions come from the so-called renormalons,
which are associated with known singularities \cite{MUELLER}
in the Borel transform
\bel{eq:borel_expansion}
\cB[D]\equiv \tilde D(b) = {1\over 4\pi^2}\sum_{n=0} K_{n+1} {b^n\over n!}
\ee
of 
$D(a)=\sum_n K_n a^n/(4\pi^2)$,
where $a\equiv\alpha_s/\pi$.
In terms of Feynman graphs, the renormalons
correspond to inserting a chain of quark-loop bubbles into a gluonic line.
Diagrams of this kind generate a factorial growth of the perturbative
$D(a)$ series (i.e. $K_n\sim n!$) at large orders.
Owing to the explicit $1/n!$ factor in the coefficients of
(\ref{eq:borel_expansion}), $\tilde D(b)$ is a much better behaved series.
Thus, one could hope to define $D(a)$ through its Borel sum
\bel{eq:borel_sum}
D(a) - D(0) = \int_0^\infty db \, \tilde D(b)\, \exp{(-b/a)} \, .
\ee
Unfortunately, the function $\tilde D(b)$ has singularities
at $b= m b_0 \equiv 2 m/(-\beta_1)\, $ ($m=-1,\pm2,\pm3,...$)
and therefore the integral (\ref{eq:borel_sum}) is not defined.

The singularities at $m=+2,+3,...$ (infrared renormalons) are generated
by the low-momenta behaviour of these higher-order diagrams,
which produces  contributions of the form
$D(a)\sim\sum_n n! \, (a/m b_0)^n$; i.e.
$\tilde D(b)\sim -m b_0/(b- m b_0)$ for $b< m b_0$.
The pole at $b=m b_0$ gives rise to an ambiguity when one tries
to reconstruct $D(a)$ from $\tilde D(b)$:
[$1/a\approx -\beta_1 \log{(\sqrt{s}/\Lambda)}$]
\bel{eq:ambiguity}
\delta D(a) \sim \exp{(-m b_0/a)} \sim \left(\Lambda^2/ s\right)^m \, .
\ee
These infrared  $1/s^m$ contributions
are reabsorbed \cite{MUELLER,GR:94,DA:84}
into the non-perturbative terms of the OPE.

The absence of a singularity at $m=1$ is in fact related to the lack of any
physical gauge-invariant local operator of dimension 2.
This was challenged in ref. \cite{BROWN}, where the
relation
$\cB[\mbox{\rm Im}\Pi](b) \sim
[\sin{\left(\pi b/b_0\right)}/\left(\pi b/b_0\right)] \,\tilde D(b)$
was established,
implying that either $\tilde D(b)$ has a pole corresponding to $m=1$ or
$\cB[\mbox{\rm Im}\Pi](b)$ has a zero at that point. The first possibility
(an infrared renormalon for $m=1$) would imply a $1/s$ ambiguity which
could not be reabsorbed by the OPE. This  puzzle has been
clarified by the exact calculation of $\tilde D(b)$ (both in QED \cite{BENEKE}
and in QCD \cite{LTM:94}) in the limit of a large number of quark flavours,
which shows that there is no infrared renormalon
corresponding to $m=1$, and $\cB[\mbox{\rm Im}\Pi](b)$
has in fact a zero.
This confirms the relation of infrared renormalons to gauge-invariant
operators \cite{MUELLER,GR:94,DA:84}.

The ultraviolet renormalons  ($m=-1,...$) are generated by the high-momenta
behaviour and, because of asymptotic freedom, give rise to a
Borel-summable series (they are in the negative real axis,
outside the integration region),
which makes them harmless.
They don't put any real limit   \cite{MUELLER}
to the applicability of perturbation theory.
Nevertheless, if the Borel summation is not performed,
the ultraviolet renormalons induce an intrinsic
uncertainty in the truncated perturbative series.

The factorial growth of the perturbative series is indeed dominated
by the contribution of the leading ultraviolet renormalon ($m=-1$),
$K_n\sim n!/(-b_0)^n$, which gives rise to an asymptotic series:
\beqn\label{eq:asymptotic_series}
|K_1 a| > |K_2 a^2| &\!\!\!\! > &\!\!\!\! \ldots > |K_{N-1} a^{N-1}| \sim
|K_N a^N|
\no \\
< |K_{N+1} a^{N+1}|  &\!\!\!\! < &\!\!\!\! \ldots
\eeqn
The successive terms decrease until $N\sim b_0/a$, where the minimum
value is attained, and the series explodes afterwards.
The alternating sign of the $K_n$ coefficients guarantees that the
series is Borel-summable. However, if one only considers the truncated
series at a given finite order, the accuracy is obviously limited by
the size of the minimum term \cite{ZAKHAROV}:
\beqn\label{eq:ur_accuracy}
4\pi^2\,\delta D(a) &\!\!\!\!\equiv &\!\!\!\!
|K_N a^N| \sim N! \, N^{-N}
\sim \sqrt{2\pi N}\, \e^{-N}
\no \\ &\!\!\!\! \sim &\!\!\!\!
\exp{(-b_0/a)} \sim \Lambda^2/ s \, .
\eeqn

Thus, the $m=-1$ ultraviolet renormalon seems to induce
an uncertainty in the perturbative series,
which scales like $1/s$.
This has been used \cite{ALTARELLI,AL:94} to advocate for an
{\em additional} (small)
uncertainty in $R_\tau$ proportional to $1/M_\tau^2$;
such a conclusion is {\it incorrect}, because:
\begin{enumerate}
\item As it stands, the estimate (\ref{eq:ur_accuracy}) 
depends on
the chosen renormalization scheme (the definition of $\Lambda$). A more
careful analysis \cite{BZ} shows that the ambiguity actually scales as
$A \sqrt{\alpha_s(\mu^2)}\Lambda^2 s/\mu^4$, where the scheme dependence
has been absorbed into the factor $A$. The result  (\ref{eq:ur_accuracy})
only follows if one takes $\mu^2=s$. Keeping an arbitrary renormalization
scale $\mu$, the $\Lambda^2 s/\mu^4$ term
does not contribute to
the close-contour integral (\ref{eq:circle}) defining $R_\tau$ (Cauchy
theorem).
\item One could still argue \cite{AL:94} that the improved perturbative series,
which resums higher-order logarithms through the
$A^{(n)}(\alpha_s)$ functions in Eq.~(\ref{eq:a_xi}), takes $\mu^2=s$ and thus
the $1/M_\tau^2$ ambiguity comes back.
However, it was
shown in ref. \cite{LDPa} that the scale dependence of the
resummed perturbative result is tiny; taking $\mu = \xi \sqrt{s}$, the
difference between $\xi = 1$ and $\xi=3$ is very small.
A $\Lambda^2 /(\xi^2 M_\tau^2)$ term
would instead change by a factor of 9, making such a contribution
irrelevant for $\xi\sim 3$.
\item
Eq.~(\ref{eq:ur_accuracy}) is just an
estimate of the ultimate perturbative error, defined as the size of the
minimum term. At the $M_\tau$ scale, $a\approx 0.1$ and therefore
$N\sim 2 /(-\beta_1 a)\sim 4$, which by no means is a large number; thus,
one could question the large-$N$ approximations used to obtain the
result\footnote{
Moreover, a recent estimate \protect\cite{VZ:94} of the contributions
coming from a double
renormalon chain (i.e. two dressed gluon propagators) has shown that
these higher-order effects are not suppressed, which makes the standard
ultraviolet-renormalon calculus ill-defined.}.
However, the important point is that one is in fact {\em assuming} that
$|K_3 a^3|> |K_4 a^4| > |K_N a^N|$, i.e. the so-called ultraviolet-renormalon
effect is {\em by definition} smaller than the usual perturbative error.
It only gives a crude estimate of the maximum accuracy which can be obtained
by computing higher-orders in perturbation theory.
{ It is not} a new source of uncertainty, but just a lower bound
on the {\em same} perturbative error.
\end{enumerate}

Any possible ultraviolet renormalon ambiguity is therefore already included
in the perturbative error quoted in Eq.~(\ref{eq:pert_error}).
Taking an extremely conservative attitude, one could argue that the onset
of the asymptotic behaviour of the perturbative series has been perhaps
reached already at $N=3$. Although, there is no signal of such behaviour
in the presently known terms, the possibility of a large $N=4$ contribution
cannot be excluded.
If that were the case, one should then take the size of the known $N=3$ term
(and not a rough large-$N$ approximation to it!)
as the ultimate uncertainty of the truncated perturbative calculation.
As explicitly shown in Eq.~(\ref{eq:divergent_series}),
the conservative perturbative error (\ref{eq:pert_error}) we have been adopting
\cite{OHIO} in the actual $R_\tau$ calculation, already includes
such a pathological possibility.


\subsection{Charm corrections}
\label{subsec:charm_corr}

Since the $\tau$ mass 
is below the charm-production threshold, the calculation of
$R_\tau$ has been performed in the effective QCD theory with only 3 active
quark flavours.
The decoupling of heavy quarks from the non-singlet currents \cite{AC:75}
guarantees that the leading (logarithmic) charm-quark effects can be reabsorbed
into the QCD running coupling \cite{OS:81}.
These contributions are then taken into
account, through the matching relation between the values of $\alpha_s$ in the
effective theories with 3 and 4 flavours \cite{BW:83,LRV:94}.
Nevertheless, there are additional corrections suppressed by inverse powers
of the heavy-quark mass, which a priori could be sizeable.
These effects appear first at $\cO(\alpha_s^2)$, through a virtual $c$-$\bar c$
vacuum-polarization contribution to the gluon propagator.
The charm-quark contribution is fully known at $\cO(\alpha_s^2)$
\cite{CH:93,SS:94}
and the $\cO\left[\alpha_s^3 \left(M_\tau/ m_c\right)^n\right]$
contributions with $n=1,2$ and $3$
have been recently computed \cite{LRV:94}.
The final numerical effect is very small \cite{LRV:94,CH:93,SS:94}:
\beqn\label{eq:charm}
\Delta_c &\!\!\!\!\! = &\!\!\!\!\! \left({\alpha_s\over\pi}\right)^2
{M_\tau^2\over m_c^2} \left[ {1\over 225}
\log{\left({m_c^2\over M_\tau^2}\right)}
+ {107\over 4500} + \ldots\right]
\no\\ &\!\!\!\! \approx &\!\!\!\!  0.0004 \, .
\eeqn

\subsection{Non-perturbative corrections}

The experimental data has already shown that the
total non-perturbative correction to $R_\tau$ is small
[see Eq.~(\ref{eq:delta_NP})], in agreement with the theoretical prediction.
Nevertheless, with the present experimental accuracy, inverse power
corrections introduce an uncertainty of about 1\%.

 On the theoretical side, it is clear that those corrections are very
suppressed
(either by a factor $1/M_\tau^6$ or $\alpha_s^2/M_\tau^4$); however, the
actual estimate of the leading contribution in Eq.~(\ref{eq:delta_6})
relies on phenomenological fits.
The main source of information has been up to now $e^+e^-$ data,
which only measures $\delta^{(6)}_{ij,V}$. In order to estimate the axial
contribution, one needs to assume
$\delta^{(6)}_{ij,A}/\delta^{(6)}_{ij,V}\approx -11/7$, as predicted by
factorization. While the opposite sign of the vector and axial-vector
$D=6$ contributions (helicity flip) does not depend on that approximation,
the magnitude of their ratio is changed by non-factorizable corrections.
Those non-factorizable radiative contributions have been found to be
small \cite{ACH:94} (in the $\overline{MS}$ scheme with a na\"{\i}ve
anticommuting
$\gamma_5$). Nevertheless, it would be desirable to have a direct
experimental measurement.

The preliminary fits of the $\tau$-decay hadronic distribution, given in
Table \ref{tab:moments}, result in a value of $\delta^{(6)}$ which has the
predicted magnitude but the wrong sign.
However,
given the present size of the experimental error bars and the strong
correlation between the different fit parameters, one can only conclude
that $\delta^{(6)}$ is in fact small.

{}From the difference between the measured semi-inclusive $\tau$ hadronic
widths
in Eq.~(\ref{eq:exp_r_v}),
$R_{\tau,V}- R_{\tau,A} = 0.10\pm 0.07$,
it is possible to extract the value of
$\delta^{(6)}_{V-A}$.
Note that the perturbative contributions cancel in this difference.
Taking into account the small $D=2$ and $D=4$ corrections \cite{BNP},
one gets
%
$\delta^{(6)}_{V-A} = (6\pm 5)\%$,
%
in good agreement with the theoretical estimate in Eq.~(\ref{eq:delta_6}).
However, the present errors are too large for this determination
to be significant. Clearly, better data is required.

There is another source of non-perturbative contributions,
which has been recently analyzed: instantons
\cite{NP:94,BBB:93}.
It is well-known that instantons do spoil the OPE by introducing
corrections that are power-suppressed by 9 or more inverse powers of
the momentum.
Owing to this large power suppression, the instanton contributions are
very sensitive to the momentum scale: they are either negligible,
and therefore irrelevant for practical applications, or very large,
in which case they destroy the convergence of the OPE.
Thus, instantons just provide a lower limit to the momentum scale at
which the OPE can be applied.

The existing analyses
\cite{NP:94,BBB:93} indicate that the instanton
contribution to $R_\tau$ is tiny.
Unfortunately,
it is difficult to make reliable quantitative predictions.
For $\alpha_s(M_\tau^2)=0.32$,
the instanton effect could reach the 0.2--0.3\% level \cite{BBB:93},
but the theoretical
estimates are no longer valid at larger $\alpha_s$ values.
Nevertheless, the instanton contributions to
$R_{\tau,V}$ and $R_{\tau,A}$
are predicted to be much larger 
than the total correction to  $R_\tau$ \cite{BBB:93}
(there is a cancellation between the vector and axial-vector contributions
to $R_\tau$,
similar to the one occurring in $\delta^{(6)}$).
The possible size of the instanton effect can then be bounded using
the present information on $R_{\tau,V/A}$.

The data points in Figure~\ref{fig:r_tau_m} have a very soft dependence on the
energy scale, which discards any
sizable instanton contribution above 1.2 GeV.
Moreover, a recent phenomenological analysis of
$R_{\tau,V}-R_{\tau,A}$, where the
instanton effect is supposed to be maximal, concludes \cite{KM:94}
that the instanton contribution to $R_\tau$ is smaller then 0.0005.

\subsection{Uncertainties in the running of $\alpha_s$}

The final determination of $\alpha_s(M_\tau^2)$ has to be compared with
the values of $\alpha_s$ measured at other scales, for instance
$\alpha_s(M_Z^2)$.
The correct way of running the strong coupling through the different
physical thresholds ($c$, $b$, \ldots) at $\cO(\alpha_s^3)$ was known
already twelve years ago \cite{BW:83}. One considers two different
strong couplings, $\alpha_s^{(f)}$ and $\alpha_s^{(f-1)}$,
in the effective QCD theories with $N_f$ and $N_{f-1}$ active flavours, and
imposes a series of matching conditions \cite{BW:83,LRV:94}
which guarantee that the same
physical predictions are obtained in both theories. Contrary to common belief,
the $\overline{MS}$ strong coupling is (in general) not continuous at the
matching point \cite{BW:83}. The exact point where the matching between
both theories is performed is actually irrelevant \cite{BE:94},
provided one does not
choose a crazy $\mu/m_q(\mu^2)$ ratio which would induce huge higher-order
corrections.

Taking $\alpha_s(M_\tau^2)=0.33$ as input, the resulting value of
$\alpha_s(M_Z^2)$ changes by less than 0.5\% if the $b$-threshold
matching point is varied between 2 and 40 GeV, or if
one changes the $c$-threshold between $M_\tau$ and $m_b$.
Note that the usual incorrect procedure \cite{ALTARELLI}
of requiring a continuous $\alpha_s$ largely overestimates this error.

The matching conditions take correctly into account the logarithmic
heavy-quark corrections. The physical observables get in addition
$1/m_q^n$ contributions that need to be computed.
It has been advocated \cite{ALTARELLI,AL:94}
that the uncertainty associated with these
quark-mass effects can be estimated through a {\em generous} variation
of the matching point.
This is not quite correct. Varying the matching point, one is only
testing the $\mu$-dependence of the logarithmic corrections.
Assuming that one knows the matching conditions to a sufficiently high
order, the variation of $\mu$ would not produce any numerical change
(i.e. zero error!); but heavy-quark effects would be of course still there.

In fact, within the $\overline{MS}$ scheme, the $1/m_q^n$ contributions
have nothing to do with the running of $\alpha_s$. The renormalization
group equations only involve logarithms, because $\overline{MS}$ is a
mass-independent renormalization scheme. Mass effects are of course taken into
account in the explicit calculation of each physical observable, as
discussed in Sect. \ref{subsec:charm_corr}.
The induced correction to $R_\tau$, given in Eq.~(\ref{eq:charm}),
is very small.

\section{SUMMARY}

Because of its inclusive nature,
the total hadronic width of the $\tau$ can be rigorously computed
within QCD.
One only needs to study two-point
correlation functions for the vector and axial-vector currents.
As shown in Eq. (\ref{eq:circle}),
this information is only needed in the complex plane,
away from the time-like axis; the dangerous region near the physical
cut does not contribute at all to the result, because of the phase-space
factor $(1-s/M_\tau^2)^2$.
The uncertainties of the theoretical predictions are then quite small.
Notice that the accuracy of the $R_\tau$ predictions
is much better than the corresponding estimate of
$R_{e^+e^-}(s)$ at $s=M_\tau^2$
[$R_{e^+e^-}(s)$   measures the vector spectral function on
the physical cut, where the theoretical predictions (at least at low
energies) are  more uncertain].

The ratio $R_\tau$ is very
sensitive to the value of the strong coupling, and therefore can be
used to measure $\alphM$ \cite{NP}.
This observation has triggered an ongoing effort to improve
the knowledge of $R_\tau$ from both  the experimental and the theoretical
sides.
The fact that $M_\tau$ is a quite low energy-scale (i.e. that $\alphM$ is big),
but still large enough to allow a perturbative analysis, makes
$R_\tau$ an ideal observable to determine the QCD coupling.
Moreover, since the error of $\alphmu$ shrinks as $\mu$ increases,
the good accuracy of the $R_\tau$ determination of $\alphM$ implies
a very precise value  of $\alpha_s(M_Z^2)$.

The theoretical analysis of $R_\tau$ has reached a very mature level.
Many different sources of possible perturbative and non-perturbative
contributions
have been
analyzed in detail. As shown in the previous section, the final theoretical
uncertainty is small and has been adequately taken into account in the
final $\alpha_s(M_\tau^2)$ determination in Eq.~(\ref{eq:alpha}).
(Note, however, that there are still sizeable fluctuations in the
$B_l$ measurements \cite{DU:94}, which could slightly modify the final result).

The comparison of the theoretical  predictions with the experimental
data shows
a successful and consistent picture.
The $\alphM$ determination is in excellent agreement with
the measurements at the $Z$-mass scale,
providing clear evidence of the running of
$\alpha_s$. Moreover, the analysis of the semi-inclusive components
of the  $\tau$ hadronic width, $R_{\tau,V}$, $R_{\tau,A}$
and $R_{\tau,S}$, and the invariant-mass distribution of the final
decay products
gives a nice confirmation of the reliability
of the theoretical
framework.

%
\section*{Acknowledgements}

I would like to thank
E. Braaten, F. Le Diberder and S. Narison for
a very rewarding collaboration.
This work has been supported in part by
CICYT (grant AEN-93-0234) and IVEI (grant 03-007).

%
%

\end{document}